\documentclass[twocolumn,showpacs,amsmath,amssymb,pra,superscriptaddress]{revtex4}

\usepackage{graphicx}
\usepackage{dcolumn}
\usepackage{bm}
\usepackage{textcomp}

\begin{document}
\title{Creating versatile atom traps by applying near resonant laser light in magnetic traps}

\date{\today}
\pacs{03.75.-b, 37.10.Gh, 42.55.Ye}

\author{Stephan Middelkamp}
\email[]{stephan.middelkamp@physnet.uni-hamburg.de}
\affiliation{Zentrum f\"ur Optische Quantentechnologien, Universit\"at
Hamburg, Luruper Chaussee 149, 22761 Hamburg, Germany}
\author{Michael Mayle}
\email[]{michael.mayle@physnet.uni-hamburg.de}
\affiliation{Zentrum f\"ur Optische Quantentechnologien, Universit\"at
Hamburg, Luruper Chaussee 149, 22761 Hamburg, Germany}
\author{Igor Lesanovsky}
\email[]{Igor.Lesanovsky@nottingham.ac.uk}
\affiliation{School of Physics and Astronomy, Faculty of Science, University of Nottingham, Nottingham, UK}
\author{Peter Schmelcher}
\email[]{Peter.Schmelcher@physnet.uni-hamburg.de}
\affiliation{Zentrum f\"ur Optische Quantentechnologien, Universit\"at
Hamburg, Luruper Chaussee 149, 22761 Hamburg, Germany}

\date{\today}

\begin{abstract}\label{txt:abstract}
We utilize the combination of two standard trapping techniques, a magnetic
trap and an optical trap in a Raman setup, to propose a versatile and
tunable trap for cold atoms. The created potential provides several
advantages over conventional trapping potentials. One can easily
convert the type of the trap, e.g., from a single well to a double well trap. Atoms in different internal states can be trapped in different trap types, thereby
enabling the realization of experiments with multi-component
Bose-Einstein condensates. Moreover, one can achieve variations of the trapping
potential on small length scales without the need of
microstructures. We present the potential surfaces for different setups, demonstrate their
tunability, give a semi-analytical expression for the potential, and propose experiments
which can be realized within such a trap.
\end{abstract}

\maketitle
\section{Introduction}
Trapped ultracold atomic gases serve as an ideal system to model many body systems and to investigate fundamental questions of quantum mechanics. 
A paradigm phenomenon are the so called Josephson oscillations, which were demonstrated in a double well potential 
for many atoms \cite{Albietz05,Schumm05,Hofferberth06,sewel}; more recently, even the tunneling of individual atoms through a barrier was demonstrated \cite{Kierig08}. A key ingredient 
for modelling new systems are novel ways of trapping cold atoms. A standard technique for trapping atoms of a single species is to use either static magnetic 
fields \cite{Folman,Wiemann,Fortagh,Dalfovo} or  a superposition of static and oscillating magnetic fields leading to the so-called radio-frequency dressed 
adiabatic potentials \cite{Lesanovsky06,Hofferberth06b,Lesanovsky06b}. Different species can be trapped in an optical trap making use of the so-called light 
shift \cite{Grimm}. The potential results from the intensity maximum (minimum) of a laser that is red (blue) detuned with respect to an atomic transition 
frequency. By superpositions of different laser beams and intensity configurations one can create, e.g., optical lattices or double well potentials.
In a more recent approach, two lasers in a Raman configuration were used  to trap atoms. Such a setup allows for example the creation of optical lattices 
with a reduced lattice spacing compared to standard optical lattices \cite{Zhang05}.

A combination of optical and magnetic fields for creating atom traps has been already described in several works. Even in one of the first BEC 
experiments a superposition of a far detuned laser and a magnetic trap was used to trap the atoms \cite{Davis}. In Ref.\ \cite{Deutsch97} Deutsch \textit{et al.} investigated the combination of a constant magnetic field and a state dependent optical lattice that allows for the creation of a  lattice of double well potentials. In more recent approaches, the superposition of radio frequency-fields and magnetic fields \cite{Lesanovsky06,Fernholz} or optical lattices \cite{Lundblad08}  were used for trapping atoms in tunable potentials. In this paper, we derive the potential for an atom exposed simultaneously to an inhomogeneous magnetic field in  a Ioffe-Pritchard trap like configuration and two lasers in a Raman configuration.  
In this case, the non-trivial combination of the magnetic and the laser fields cannot longer be reduced to a potential resulting from an effective magnetic field. A direct consequence of this fact is that the potential surfaces for different hyperfine components of an atom do not only differ by a global factor but can be 
substantially different. Thus, it is possible to confine the different components of a multi-component BEC in traps of different frequencies or different types, e.g., one component in a double well trap  and another one in a single well trap located at the barrier of the double well. Moreover, due to the availability of additional parameters, flexibility is gained in shaping the potentials compared to conventional traps. For example, one can smoothly convert a single well potential into a double well potential or drive a double well potential by varying an offset magnetic field. Furthermore, it is possible to rotate the potential around one axis by changing the phase between the Raman lasers.

The paper is structured as follows. In Section II, we derive the effective Hamiltonian. In Section III our numerical results are presented. Specifically, we provide an 
overview of the potential surfaces of the different components and investigate the transition of a double well to a single well potential as well as the rotation 
of a double well potential for one component in detail. In Section IV we derive a semi-analytical expression for the potential surfaces; in Sec.~V 
loss mechanisms are discussed. Finally, in Sec.~ VI we summarize our results and mention directions for possible future studies.  
%
\section{Analytical Considerations}
\subsection{Hamiltonian and Setup}
The Hamiltonian of an (alkali) atom simultaneously exposed to magnetic and laser fields reads
\begin{equation}\label{hini}
H=-\frac{\hbar^2}{2M}\mathbf{\nabla_{\mathbf{R}}}^2 + H^{\text{e}}(\mathbf{r})+V^{\text{IP}}(\mathbf{r},\mathbf{R}) +V^{\text{AF}}(\mathbf{r},\mathbf{R}).
\end{equation}
Here, $\mathbf{R}$ denotes the center of mass coordinate of the atom and $\mathbf{r}$ the coordinate of the valence electron relative to the center of 
mass position; $M$ is the total mass of the atom. $H^\text{e}(\mathbf{r})$ accounts for the field-free electronic structure of the atom; for the scope of 
this work, we use $^{87}$Rb as a paradigm. $V^\text{AF}(\mathbf{r,R})$ and $V^\text{IP}(\mathbf{r,R})$ denote the contributions of the Raman lasers and 
the magnetic field, respectively.
In order to solve the coupled Schr\"odinger equation associated with Hamiltonian (\ref{hini}), we employ a Born-Oppenheimer separation of the center of 
mass motion and the electronic degrees of freedom. We are thereby led to an effective electronic Hamiltonian that parametrically depends on the center of 
mass position, 
\begin{equation}
H^{\text{eff}}(\mathbf{R})= H^{\text{e}}(\mathbf{r})+V^{\text{IP}}(\mathbf{r};\mathbf{R}) +V^{\text{AF}}(\mathbf{r};\mathbf{R}).
\label{effective Hamiltonian}
\end{equation}
Its solutions $V_\kappa(\mathbf R)$ serve as adiabatic potential energy surfaces for the center of mass motion of the atom; each of these (trapping)
potentials is then associated with a given internal state $\kappa$ of the atom.

Regarding the magnetic field configuration, we consider the setup of a Ioffe-Pritchard trap \cite{Pritchard} which is given by a two-dimensional quadrupole field in the $x_1,x_2$-plane together with a perpendicular offset (Ioffe-) field in the $x_3$-direction; it can be parameterized as  $\mathbf B(\mathbf x)=Gx_1\mathbf{e}_1-Gx_2\mathbf{e}_2+B_I\mathbf{e}_3$. $G$ denotes the magnetic gradient of the two-dimensional quadrupole field and $B_I$ the  constant offset field oriented along the $x_3$-axis. The quadratic term $\mathbf B_q\propto (x_3^2-\rho^2/2)\mathbf e_3$ 
that usually arises for a Ioffe-Pritchard configuration can be exactly zeroed by geometry, which we are considering in the following. In actual experimental 
setups, $\mathbf B_q$ provides a weak confinement also in the $x_3$-direction. Omitting $\mathbf B_q$, the magnitude of the magnetic field at a certain 
position $\mathbf x$ in space is given by $|\mathbf B(\mathbf x)|=\sqrt{B^2+G^2\rho^2}$, which yields a linear asymptote 
$|\mathbf B(\mathbf x)|\rightarrow G\rho$ for large values of the coordinates ($\rho=\sqrt{x_1^2+x_2^2}\gg B/G$) and a harmonic behavior 
$|\mathbf B(\mathbf x)|\approx B+\frac{1}{2}\frac{G^2}{B}\rho^2$ close to  the origin ($\rho\ll B/G$).
The magnetic field interaction within the Born-Oppenheimer approximation reads
\begin{equation}
V^\text{IP}(\mathbf{r;R})=g_F\mu_B \mathbf{F}\cdot \mathbf{B(R)}. 
\end{equation}

\begin{figure}
 \includegraphics[width=8.5cm]{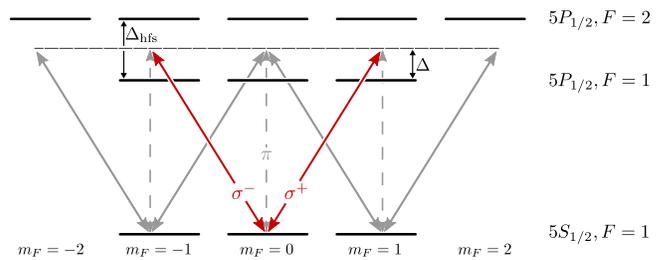}  
  \caption{State linkage diagram of the unperturbed atom. The solid arrows denote the transitions allowed for $\sigma^+$ and $\sigma^-$ light, whereas  the dashed arrows indicate allowed transitions with $\pi$ polarized light. The energy gap between the $5P_{1/2},F=1$ and $5P_{1/2},F=2$ manifold is $\Delta_{\text{hfs}}= 2\pi \hbar \times 0.8$ GHz \cite{Arimondo}.}
  \label{fig: fig1}
\end{figure}
\begin{figure}
 \includegraphics[width=8.5cm]{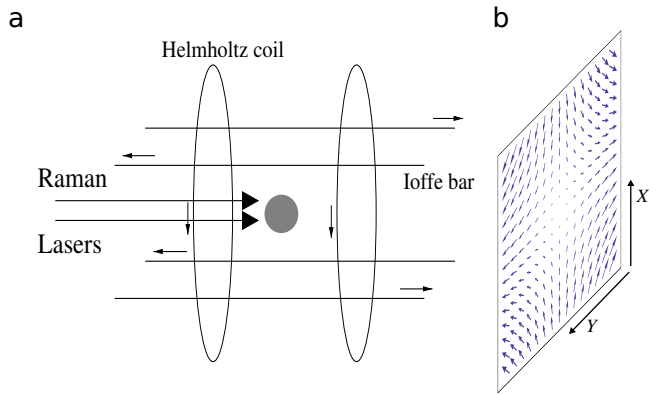}  
  \caption{(a) Setup showing the propagation direction of the laser beams (large arrows) and the configuration of the magnetic trap. The Helmholtz coils generate the homogeneous Ioffe field oriented along the $Z$ direction and the Ioffe bars the quadrupole field in the $X-Y$ plane, shown in subfigure (b).}
  \label{fig: fig2}
\end{figure}

The Raman configuration of the excitation lasers is depicted in Fig.~\ref{fig: fig1}. It consists of two oppositely circular polarized lasers that are 
close to resonance with the $D_1$ transition line, i.e., being blue-detuned by $\Delta$ with respect to the transition from the $5S_{1/2},F=1$ ground 
state manifold to the  $5P_{1/2},F=1$ excited state of $^{87}$Rb. The propagation direction of the lasers is chosen to coincide with the direction of the constant 
Ioffe Field $B_I$. The overall setup is shown in Fig.\ \ref{fig: fig2}

 Within the dipole approximation, the potential of an atom exposed to the laser fields is given by 
\begin{equation}
 V^{\text{AF}}(\mathbf{r};\mathbf{R})=-e\mathbf{r}\cdot\mathbf{E}(\mathbf{R},t),
\end{equation}
$\mathbf{E}(\mathbf{R},t)$ denoting the electric field of the lasers. The latter can be expressed as
\begin{equation}
 \mathbf{E}(\mathbf{R},t)=\frac{1}{2}\sum_{i=1}^2 \left[\mathbf{E}_i(\mathbf{R},t)+\mathbf E^{\star}_i(\mathbf{R},t)\right]
\end{equation}
with $\mathbf{E}_i(\mathbf{R},t)=\boldsymbol{\epsilon}_i\varepsilon_i(\mathbf{R})e^{-i(\mathbf{k}_i\mathbf{R}-\omega_it+\phi_i(t))}$ being the electric 
field associated with the $i$th laser. The amplitudes $\varepsilon_i(\mathbf{R})$ of the electric fields  are spatially dependent in order to account for the 
focussing and shape of the laser beams. The factors $\boldsymbol{\epsilon}_i$ are the unit polarization vectors given by 
$\boldsymbol{\epsilon}_1=\frac{1}{\sqrt{2}}(\mathbf{e}_1+i\mathbf{e}_2)$ ($\sigma^+$ light) and  
$\boldsymbol{\epsilon}_2=\frac{1}{\sqrt{2}}(\mathbf{e}_1-i\mathbf{e}_2)$ ($\sigma^-$ light), respectively.
$\phi_i(t)$ take into account the phases of the lasers, which additionally can depend on time. 
Expanding the atom field interaction in the basis $|\alpha\rangle$, defined as the eigenfunctions of the field free atom, leads to
\begin{align}
 V^{\text{AF}}(\mathbf{r};\mathbf{R})={}&\frac{1}{2}\sum_{i=1}^2\sum_{\alpha,\gamma}(\omega^{(+)}_{i,\alpha \gamma}
+\omega^{(-)}_{i,\alpha \gamma})|\gamma\rangle\langle\alpha|+\text{h.c.}
\end{align}
with $\omega^{(+)}_{i,\alpha \gamma}=e\langle\alpha|\mathbf{E}_i(t)\mathbf{r}|\gamma\rangle$ and 
$\omega^{(-)}_{i,\alpha \gamma}=e\langle\alpha|\mathbf{E}^{\star}_i(t)\mathbf{r}|\gamma\rangle$. 

%
\subsection{Rotating Wave Approximation}
We employ the rotating wave approximation \cite{scully} in order to remove the time-dependence of Hamiltonian (\ref{effective Hamiltonian}) that arises due to the laser interaction. For reasons of simplicity we assume that both lasers have the same frequency $\omega_i\equiv\omega$ and the 
same profile $\varepsilon_i(\mathbf{R})\equiv\varepsilon(\mathbf{R})$. Since the magnetic field interaction term is block diagonal, i.e., does not mix states with different 
total angular momenta, $V^{\text{IP}}(\mathbf{r;R})$ is not affected by the transformation into the rotated frame. The transformed Hamiltonian reads
\begin{equation}\label{eq: Hamiltonian RWA}
H_{\text{RWA}} = V^{\text{IP}}+V^{\text{e}}_{\text{RWA}}+V^{\text{AF}}_{\text{RWA}}, 
\end{equation}
where
\begin{align}
V^{\text{e}}_{\text{RWA}}&=\sum_\alpha E_\alpha |\alpha\rangle\langle\alpha | + \sum_l (E_l-\hbar \omega) |l \rangle\langle l| \\
V^{\text{AF}}_{\text{RWA}}&=\frac{1}{2}\sum_{i=1}^2 \sum_{\alpha,l} \varepsilon(\mathbf{R}) e^{-i(\mathbf{k}_i\mathbf{R}+\phi_i(t))} \langle l | \boldsymbol{\epsilon}_i \mathbf r|\alpha\rangle |l\rangle \langle \alpha| +\text{h.c.}
\end{align}
(for clarity, we omit the arguments $\mathbf r$ and $\mathbf R$ of the potentials in the following).
Here, $\alpha$ labels the different states of the ground state manifold and $l$ labels the excited states. Since the lasers are close to resonance 
to the $D_1$ transition line, we can restrict our basis to states close to the ground state and the first excited state. Using as a basis all hyperfine states 
of the $5S$ and $5P$ manifolds leads to an effective $32 \times 32$ matrix that will be diagonalized.
%
%

%
\subsection{Van Vleck Perturbation Theory}
In the last subsection we derived the Hamiltonian expanded in the eigenfunctions of the unperturbed atom. One can use 
\emph{van Vleck perturbation theory} \cite{Shavitt} to adiabatically eliminate the excited $5P$ levels that serve as intermediate states for the Raman 
transitions. In this manner, the Hamiltonian can be reduced to an operator acting only on the ground state manifold 
(i.e., all states $|\alpha\rangle=|5S_{1/2},F=1,m_F\rangle$ with $m_F\in\{0,\pm1\}$),
\begin{equation}\label{eq: Hamiltonian VV}
 H_{\text{VV}}=V^{\text{IP}}+\sum_\alpha E_\alpha |\alpha\rangle\langle\alpha |+\sum_{\alpha,\beta} \mathcal W_{\beta \alpha}|\beta\rangle\langle\alpha|,
\end{equation}
with
\begin{eqnarray}\label{eq:dressWba}
\mathcal W_{\beta\alpha}&=&\frac{1}{2}\sum_l\mathcal V_{\beta l}\mathcal V_{l\alpha}\Big(\frac{1}{E_\alpha-E_l}+\frac{1}{E_\beta-E_l}\Big)\,,
\end{eqnarray}
being the effective interaction within the ground state manifold. Here, the index $l$ labels the excited states, which have been eliminated. For a detailed 
derivation of Eq.~(\ref{eq:dressWba}), we refer the reader to the appendix of Ref.~\cite{mayle10}. Employing
\begin{equation}
\mathcal V_{l\alpha}=\frac{1}{2}\sum_{i=1}^2\sum_{\alpha,l} \varepsilon(\mathbf{R}) e^{-i(\mathbf{k}_i\mathbf{R}+\phi_i(t))} \langle l | \boldsymbol{\epsilon}_i \mathbf r|\alpha\rangle.
\end{equation}
for the block off-diagonal matrix elements of $V_\text{RWA}^\text{AF}$ yields the effective interaction $W$ within the ground state manifold 
whose matrix representation correspondingly reads
\begin{align}\label{Weffective}
W_{\beta \alpha}={}&\frac{1}{8}\varepsilon(\mathbf{R})^2 \sum_{i,i'=1}^2\sum_l e^{i[\phi_i(t)-\phi_{i'}(t)]}\langle \beta | \boldsymbol{\epsilon}_i \mathbf r|\ l\rangle \langle l | \boldsymbol{\epsilon}_{i'} \mathbf r|\alpha\rangle\nonumber\\
&\times(\frac{1}{E_\alpha-E_l+\hbar \omega}+\frac{1}{E_\beta-E_l+\hbar \omega}).
\end{align}

If one restricts the sum over the intermediate states to the $5P_{1/2}$ hyperfine sublevels, which represent a good approximation, one obtains the compact form
\begin{equation}\label{Wmatrix}
\mathcal W=\frac{|\langle 5S_{1/2}||er||5P_{1/2}\rangle|^2}{72 c\epsilon_0 \Delta}
\begin{pmatrix}
      A             & 0 & C  \\
		0             & B & 0   \\
		C^{\star}     & 0 & A 
\end{pmatrix}
\end{equation}
for the atom laser interaction, with
\begin{align}
A &= \left(1+\frac{7}{1-\frac{\Delta_{\text{hfs}}}{\Delta}}\right)I(\mathbf{R}),\\
B &= 2\left(1+\frac{3}{1-\frac{\Delta_{\text{hfs}}}{\Delta}}\right)I(\mathbf R),\\
C &= -\left(1-\frac{1}{1-\frac{\Delta_{\text{hfs}}}{\Delta}} \right)e^{i \Delta\phi(\mathbf{t})}I(\mathbf R),
\end{align}
$\Delta=E_{5S_{1/2},F=1}-E_{5P_{1/2},F=1}-\hbar \omega$ being the detuning of the transition lasers and $\Delta \Phi(t)=\phi_1(t)-\phi_2(t)$ their phase 
difference. $I(\mathbf{R})=c\epsilon_0|\varepsilon(\mathbf R)|^2/2$ denotes the intensity of the laser, $\epsilon_0$ being the dielectric constant and $c$ the speed of 
light. The reduced matrix element in Eq.~(\ref{Wmatrix}) can be deduced from the measured lifetime of the excited state which yields in our case 
$\langle 5S_{1/2}\Vert er\Vert5P_{1/2}\rangle=2.9919 \,ea_0$ \cite{loudon,volz96}.
The individual contributions of the matrix (\ref{Wmatrix}) can be interpreted as follows. The diagonal elements stem from the light shift potential of the 
lasers, i.e., the off-resonant coupling of a $m_F$ component of the ground state to an excited state. The off-diagonal elements arise due to the coupling of the $m_F=1$ ($m_F=-1$) component via an intermediate (excited) state to the  $m_F=-1$ ($m_F=1$) 
component. These off-diagonal matrix elements are not present in radio frequency traps. The specific form of the matrix occurs since the laser light is circularly polarized. Other polarizations would lead to a coupling of different states, i.e., different off-diagonal entries of the matrix $\mathcal W$.
Furthermore, we should note at this point that the matrix is expanded in the field-free basis of the atom. In this basis, the contribution of the magnetic 
field interaction $V^\text{IP}$ for the ground state manifold is represented by the spin matrices for total spin $F=1$, giving rise to off-diagonal matrix 
elements as well. As a result, the combined action of the magnetic and laser fields leads to the mutual coupling of all magnetic sublevels of the 
$5S_{1/2},F=1$ manifold, represented by a fully occupied matrix. In Section \ref{Semi-analytical} we tackle this issue by performing a principal axis 
transformation that diagonalizes the magnetic field interaction and thus provides us with a more suitable basis for the interpretation of the underlying physics.

\section{Numerical Results}\label{sec:numerical}
We restrict our investigations to both lasers having a Gaussian profile with a width $\sigma$ in the $X,Y$ plane 
and assume that they have a constant intensity in the propagation direction $Z$,
\begin{equation}
 I(\mathbf R)=I_0 \exp(-\frac{X^2+Y^2}{\sigma^2}).
\end{equation}
Since the magnetic field interaction term is independent of $Z$ as well, we find a total potential that is constant in the $Z$-direction.
In order to provide a confinement in the $Z$-direction one can, e.g., utilize an additional laser or make use of the defocusing of the laser beams. 
If not stated otherwise, we fix the detuning of the lasers to $\Delta=-\Delta_{\text{hfs}}/2$, i.e., right in the middle between the $5P_{1/2},F=1$ 
and the $5P_{1/2},F=2$ excited states as depicted in Fig.~\ref{fig: fig1}, and the intensity to $I_0=10$ W/m$^2$.

\subsection{Overview Over All Components}
\begin{figure}
\includegraphics[width=8cm]{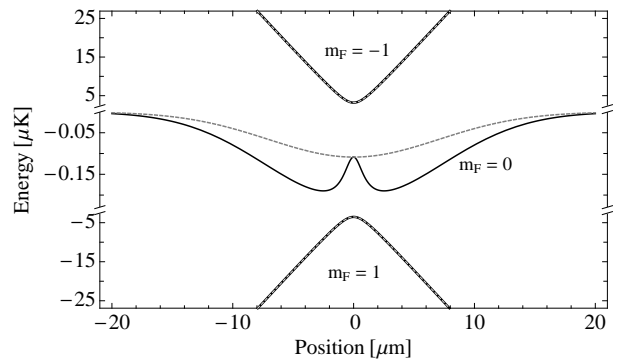}  
 \caption{Potential curves for $Y=0$ (solid line) and for $X=0$ (dashed line). Two components show an attractive potential in two dimensions whereas one component is exposed to a repulsive potential. The individual potential surfaces are well separated.}
  \label{fig: fig3}
\end{figure}
Diagonalizing the Hamiltonian matrix (\ref{eq: Hamiltonian VV}) leads to the adiabatic potential surfaces $V_\kappa(\mathbf R)$ for the center of mass motion as a function of the center of mass coordinate $\mathbf R$.
As expected from the magnetic field interaction, we find one trapped and one anti-trapped component, according to the quantum numbers $m_F=+1$ and $m_F=-1$, respectively. Interestingly, the $m_F=0$ component -- that is untrapped in a pure Ioffe-Pritchard field -- now shows an attractive potential with a double-well structure along the $X$-axis.
Note that the quantum number $m_F$ is only valid in a rotated frame of reference that we are going to introduce in Sec.~\ref{Semi-analytical}.
Nevertheless, we continue to use $m_F$ as a label for the different states in the laboratory frame in order to avoid confusion. Since the different potential surfaces are separated, one can uniquely assign to each state one particular surface.

\begin{figure}
\includegraphics[width=7.5cm]{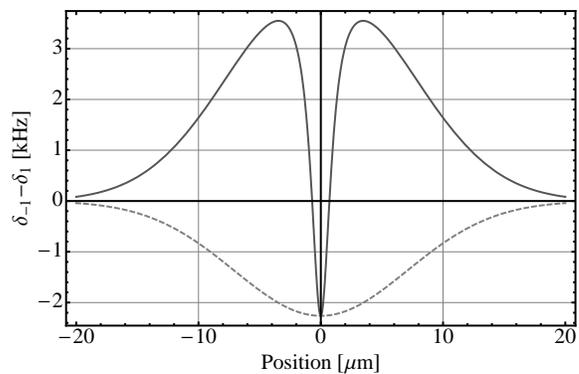}  
 \caption{Difference of the potential differences for adjacent components for $X=0$ (solid line) and $Y=0$ (dashed line). The deviation between the transition frequencies of adjacent potential surfaces is non-zero and depends on the center of mass position of the atom.}
  \label{fig: fig4}
\end{figure}
For investigating the energy spectrum further, let us define the radio transition frequencies $\delta_{-1}(\mathbf R)=V_{m_F=-1}(\mathbf R)-V_{m_F=0}(\mathbf R)$ and $\delta_{1}(\mathbf R)=V_{m_F=0}(\mathbf R)-V_{m_F=1}(\mathbf R)$ between the $m_F=0$ and the $m_F=\mp1$ states, respectively. Figure \ref{fig: fig4} provides a measure for the deviation of both transition frequencies by showing the difference $\delta_{-1}(\mathbf R)-\delta_{1}(\mathbf R)$ for $X=0$ (solid line) and $Y=0$ (dashed line). The fact that the transition frequencies do not coincide, i.e., the difference being non-zero, can be used to mutually couple two components without coupling to the third component. This allows for example a transfer of atoms from the $m_F=-1$ component to the $m_F=0$ component without coupling to the untrapped $m_F=1$ component. Moreover, the radio transition frequencies depend on the center of mass position $\mathbf R$ and thus on the absolute value of the potentials. Therefore, one may couple energy-selectively one component to another one, i.e., couple atoms of the $m_F=0$ or $m_F=-1$ component at a certain position to the untrapped $m_F=1$ component. Such a scheme can be used to evaporatively cool the $m_F=0$ or the $m_F=-1$ component \cite{Alzar06}.

%
\subsection{$m_F=0$ Component}
\begin{figure*}
\includegraphics[width=17cm]{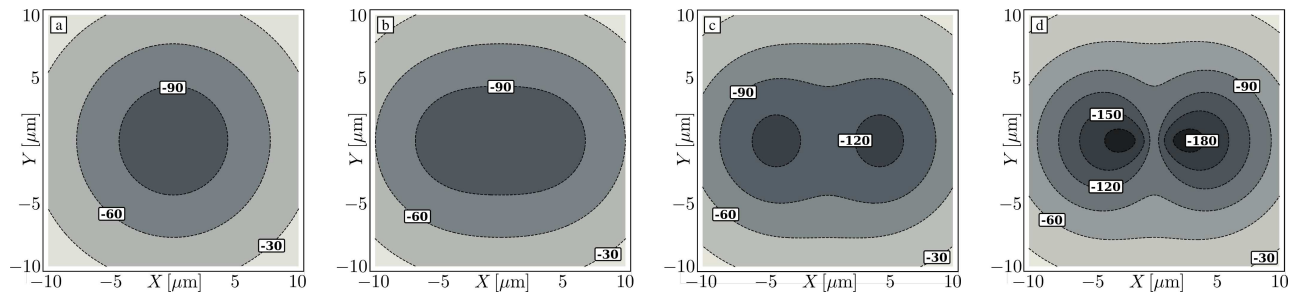}
  \caption{Contour plot of the potential surface for the $m_F=0$ component for $\Delta \Phi=0$, $\sigma=10$ $\mu$m, $G=0.1 $G/$\mu$m, and 
(a) $B_I=10$ G, (b) $B_I=1$ G, (c) $B_I=0.5$ G, and (d) $B_I=0.1$ G. The grey coded values of the potential are given in nK. By modulating the magnitude of the Ioffe field one can transform the potential smoothly from a single well to a double well potential.}
  \label{fig: fig5}
\end{figure*}
We start our detailed investigations of the individual components with the potential surface for the $m_F=0$ component. Note that this component is not confined in a pure magnetic trap. Figure \ref{fig: fig5} shows the contour plot of its trapping potential for $\Delta \Phi=0$, $\sigma=10$ $\mu$m, $G=0.1 $G/$\mu$m, and 
(a) $B_I=10$ G, (b) $B_I=1$ G, (c) $B_I=0.5$ G, and (d) $B_I=0.1$ G. The shape of the potential correspondingly changes from (a) close to being rotationally symmetric to (b) a cigar shaped potential in the $X$-direction, and finally to (c,d) a double well in the $X$-direction. 
Thus, one can change the shape of the potential from a single- to a double well potential by changing the magnitude of the Ioffe field.
Alternatively, one may also drive the double or single well potential by modulating the magnitude of the Ioffe field. 
The parameters of the double well trap can be tuned in different ways. The height of the barrier can be tuned by changing the ratio $\xi=G/B$. This is shown in detail in Fig. \ref{fig: fig8} in Sec.~\ref{Semi-analytical}.
The position of the minima can be controlled by changing the width of the lasers. This leads at the same time 
to a change of the height of the barrier and consequently to a change in the number of trapped states within each well.
For a more detailed discussion of the properties of the double well potential, we refer the reader to Sec.~\ref{Semi-analytical} where a semi-analytical expression for the potential surface is derived. By displacing the center of the laser beams in the $X$-direction with respect to the Ioffe Pritchard trap, one can create in addition a tilted double well potential.

%
\begin{figure}
\includegraphics[width=8.5cm]{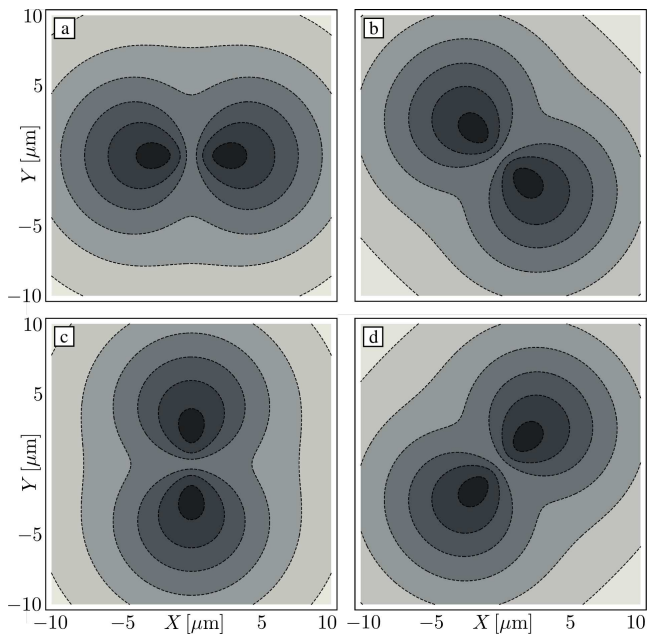}
 \caption{Contour plot of the potential surface for the $m_F=0$ component for $G=0.1$ G/$\mu$m, $\sigma=10$ $\mu$m, $B_I=0.1$ G, and for (a) $\Delta \Phi=0$, (b) $\Delta \Phi=\pi/2$, (c) $\Delta \Phi=\pi$, and (d) $\Delta \Phi=3\pi/2$. A phase difference of $\Delta \Phi$ between the Raman lasers leads to rotation of  $\Delta \Phi/2$ of the potential surface around the $Z$-axis.}
  \label{fig: fig6}
\end{figure}
Figure \ref{fig: fig6} shows the effects of a phase difference between the two excitation lasers on the potential surface. A phase difference $\Delta \Phi$ 
leads to a rotation of the whole potential surface about the $Z$-axis by $\Delta \Phi/2$. For a zero phase difference one can add the 
electric fields of the lasers, resulting in an effective electric field which is polarized linearly along the $X$-axis, whereas a phase difference of
$\pi$ leads to an effective electric field which is polarized linearly along the $Y$-axis. In general, a phase difference of $\Delta \Phi$ leads to a 
rotation of the polarization vector of the total electric field by $\Delta \Phi/2$. 
The sensitivity of the alignment of the double well potential on the orientation of the polarization vector seems at first glance surprising, given the azimuthal symmetry of a pure Ioffe-Pritchard trap. It is rooted in the spatially varying quantization axis of the Ioffe-Pritchard trap.
This issue is analyzed further in Sec.~\ref{Semi-analytical}, to which we refer the reader at this point.

%

\begin{figure}
\includegraphics[width=5cm]{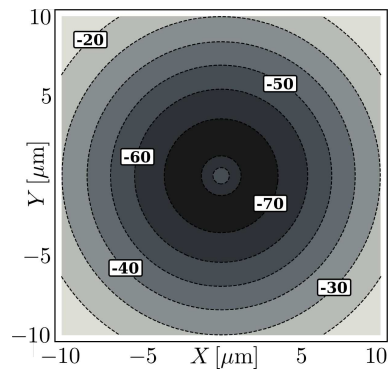}  
   \caption{Potential surface for the case of a single $\sigma^-$ polarized laser for $\sigma=10\mu$ m, $G=0.1$ G/$\mu$m, and $B_I=0.1$ G. The potential is ring shaped.}
  \label{fig: fig7}
\end{figure}

Figure \ref{fig: fig7} shows the potential surface for the same setup as in Fig.~\ref{fig: fig6} but for the case of a single $\sigma^-$-polarized laser instead of a pair of $\sigma^+ / \sigma^-$ polarized lasers. The potential is rotationally symmetric with a local maximum at the origin. Therefore it can be used as a ring shaped trap. The absolute value of the potential scales with the laser intensity. Hence, one can increase the height of the barrier by increasing the laser intensity. The position of the local minimum can be varied by changing the ratio $B/G$ or the width of the laser $\sigma$. 

%
\subsection{$m_F=-1$ component}
The potential of the $m_F=-1$ component is dominated by the contribution of the magnetic field and therefore resembles the attractive potential of a pure Ioffe-Pritchard trap, namely a single well with a minimum at the origin. Without lasers, the confinement in the $X$- and $Y$-directions is equal, leading to an isotropic potential. The contributions of the lasers break this symmetry, giving rise to a slightly ellipsoidal potential. The value of the eccentricity depends on the intensity of the lasers. A phase difference between the lasers leads to a rotation of the (anisotropic) potential surface about the $Z$-axis.

%
\section{Semi-analytical Interpretation of the Potential} \label{Semi-analytical}
\subsection{Principal Axis Transformation}
In the previous sections we investigated the system in the laboratory frame of reference where the quantization axis for the atom is determined by the direction of the constant Ioffe field. However, because of the inhomogeneity of the magnetic field, a more adequate description of our system is to define the quantization axis along the \emph{local} magnetic field vector $\mathbf{B(R)}$. In this chapter, we tackle this issue by introducing the spatially dependent unitary transformation 
\begin{equation}
 U_r=\exp(-i\alpha F_x)\exp(-i\beta F_y)
\end{equation}
that rotates the local magnetic field vector into the $z$-direction of the laboratory frame of reference with the total spin vector $\mathbf{F}=\mathbf{L}+\mathbf{S}+\mathbf{I}$ consisting of the sum of the electronic orbital angular momentum vector $\mathbf{L}$, the electronic spin vector $\mathbf{S}$, and the nuclear spin vector $\mathbf{I}$. The corresponding rotation angles are defined by
$\sin\alpha\!=\!-GY/\sqrt{B^2+G^2(X^2+Y^2)}$,
$\sin\beta\!=\!GX/\sqrt{B^2+G^2X^2}$,
$\cos\alpha\!=\!\sqrt{B^2+G^2X^2}/\sqrt{B^2+G^2(X^2+Y^2)}$, and
$\cos\beta\!=\!B/\sqrt{B^2+G^2X^2}$.
This rotation diagonalizes the magnetic field contribution in Hamiltonian (\ref{effective Hamiltonian}), giving rise to
\begin{equation}
 U_rV^\text{IP}U_r^\dagger=g_F\mu_B F_z |\mathbf{B(R)}|.
\end{equation}
Note that in the rotated frame of reference without lasers $m_F$ remains a good quantum number even in the presence of the inhomogeneous Ioffe-Pritchard field. In absence of the Raman lasers, the trapping potentials correspondingly read $V_\kappa=g_F\mu_B m_F |\mathbf{B(R)}|$.

In order to solve the time-dependent Schr\"odinger equation associated with Hamiltonian (\ref{effective Hamiltonian}), the Hamiltonian for the atom in the Ioffe-Pritchard trap and the laser interaction must be expressed in the same frame of reference. Hence, the unitary transformation $U_r$ must be applied to $V^\text{AF}$ as well. We find
\begin{equation}
 U_r\mathbf rU_r^\dagger=
\left(\begin{array}{c}
x\cos\beta + y\sin\alpha\sin\beta - z\cos\alpha\sin\beta\\
y\cos\alpha + z\sin\alpha\\
x\sin\beta - y\sin\alpha\cos\beta + z\cos\alpha\cos\beta
\end{array}\right)\,.
\end{equation}
Therefore, the $\sigma^+$ and $\sigma^-$ laser transitions that are depicted in Fig.~\ref{fig: fig1} become
\begin{align}
 \boldsymbol{\epsilon}_\pm\cdot U_r\mathbf rU_r^\dagger={}&\frac{1}{\sqrt{2}}\big[x\cos\beta+y\sin\alpha\sin\beta\nonumber\\
&\quad\,\,-z\cos\alpha\sin\beta
\pm i(y\cos\alpha+z\sin\alpha)\big]\,.\label{eq:dressUeps}
\end{align}
Equation (\ref{eq:dressUeps}) can be rewritten in terms of the polarization vectors $\tilde{\boldsymbol{\epsilon}}_\pm$ and $\tilde{\boldsymbol{\epsilon}}_0$ defined in the rotated frame of reference, showing that in a Ioffe-Pritchard trap contributions of all polarizations emerge away from the trap center \cite{mayle09,mayle10}. This changes drastically the simple transition scheme caused by the $\sigma^+$ and $\sigma^-$ light as depicted in Fig.~\ref{fig: fig1}, leading to a spatially dependent coupling between the involved ground- and excited states. 
%
\subsection{Effective potential for the $m_F=0$ component}\label{sec: effective potential}
For our setup as depicted in Fig.~\ref{fig: fig1} and zero relative phase $\Delta \Phi=0$, the operator describing the interaction between the laser field and the atom reads in the rotated frame of reference
\begin{align}
 \hat O(X,Y)\equiv{}&(\boldsymbol{\epsilon}_+ + \boldsymbol{\epsilon}_-)\cdot U_r\mathbf rU_r^\dagger\nonumber\\
={}&\sqrt{2}x\cos\beta + \sqrt{2}y\sin \alpha \sin \beta - \sqrt{2}z\cos \alpha \sin \beta.
\end{align}
Note that the dependence of $\hat O(X,Y)$ on the center of mass coordinates $X$ and $Y$ stems from the dependence of the rotation angles $\alpha$ and $\beta$ on these coordinates.
One finds the following limits. At the origin, the transformation is given by unity, providing $\hat O(0,0)=\sqrt{2}x$, i.e., $\pi$-polarized light 
in the $x$-direction. For $X=0$, $\hat O(X,Y)$ is invariant under $U_r$ which results in $\hat O(0,Y)=\hat O(0,0)=\sqrt{2}x$.
For $Y=0$ and $X\rightarrow\infty$ the gradient field dominates and the operator corresponds to the operator of $\pi$-polarized light in the $z$-direction $\hat O(\infty,0)=\sqrt{2}z$. 
A non-zero relative phase between the lasers leads to a different polarization of the total electric field in the laboratory frame, giving rise to a rotation of the operators in the rotated frame. This explains the rotation of the potential as seen in Fig.~\ref{fig: fig6}. Since it is straightforward to generalize our results to non-zero relative phases, we will restrict our analytical considerations to a zero relative phase in the following.

Within van Vleck perturbation theory as introduced in Sec.~\ref{effective Hamiltonian}, the effective interaction in the rotated frame of reference becomes
\begin{align}
\mathcal W_{\beta \alpha}={}&\frac{1}{8}\varepsilon(\mathbf{R})^2\sum_{i,i'=1}^2\sum_l \langle \beta | \boldsymbol{\epsilon}_i U_r\mathbf rU_r^\dagger|l\rangle \langle l | \boldsymbol{\epsilon}_i U_r\mathbf rU_r^\dagger|\alpha\rangle\nonumber\\
&\quad\times(\frac{1}{E_\alpha-E_l+\hbar \omega}+\frac{1}{E_\beta-E_l+\hbar \omega}), 
\end{align}
cf.\ Eq.~\ref{Weffective}. Note, that we assumed once more that the shapes and frequencies of both lasers are identical. 
We are interested in the regime of large enough magnetic fields (adjustable by the homogeneous Ioffe field component $B_I$) where the Zeeman splitting overcomes the light shifts of the laser fields, i.e., $|\mathcal W_{\beta \alpha}|\ll g_F\mu_B |\mathbf{B(R)}|$. Hence, we can approximate the fully occupied effective interaction matrix $\mathcal V^\text{IP}+\mathcal W$ by omitting the off-diagonal matrix elements $\mathcal W_{\beta \alpha},\alpha\neq\beta$, that couple the Zeeman-splitted $m_F$ components. This procedure leaves us with the diagonal matrix elements of the effective interaction,
\begin{align}
\mathcal W_{\alpha \alpha}={}&\frac{1}{2}\varepsilon(\mathbf{R})^2\sum_l\Big[
\cos^2\beta|\langle \alpha | x|l\rangle|^2\nonumber\\
&\quad+\sin^2\alpha\sin^2\beta|\langle \alpha | y| l\rangle|^2\nonumber\\
&\quad+\cos^2\alpha\sin^2\beta|\langle \alpha | z| l\rangle|^2\Big]/(E_\alpha-E_l+\hbar\omega).
\end{align}
Employing $|\langle \alpha | x| l\rangle|^2=|\langle \alpha | y| l\rangle|^2$ and performing the sum over all intermediate states $|l\rangle$ eventually yields for the $m_F=0$ component the effective potential
\begin{align}\label{eq: V_eff}
V_\text{eff}=V_0+ \frac{X^2}{\xi^2+X^2+Y^2}(V_\infty-V_0),
\end{align}
where
\begin{align}
 V_0={}&-\frac{I(\mathbf R)}{36c\epsilon_0}\left(\frac{1}{\Delta}+\frac{3}{\Delta+\Delta_{\text{hfs}}}\right)|\langle 5S_{1/2}||er||5P_{1/2}\rangle|^2,\\
 V_\infty={}&-\frac{I(\mathbf R)}{9c\epsilon_0}\frac{|\langle 5S_{1/2}||er||5P_{1/2}\rangle|^2}{\Delta+\Delta_{\text{hfs}}}
\end{align}
are the light shifts at the origin and in the limit $Y=0,X\rightarrow\infty$, respectively. $\xi=B/G$ is a length scale characterizing the particular configuration of the Ioffe-Pritchard trap.
Numerical comparison of the effective potential (\ref{eq: V_eff}) with the corresponding eigenvalue of the full  problem (\ref{eq: Hamiltonian RWA}) shows a excellent agreement for small laser intensities. We observe a maximal relative deviation of less then $1$\textperthousand{} for $I=10$ W/m$^2$. For larger intensities, the agreement gets worse since the off-diagonal matrix elements $\mathcal W_{\beta\alpha}$ increase in magnitude. However, even for $I=100$ W/m$^2$ the deviation is less than  $5$\textperthousand.
%
\subsection{Discussion of the Effective Potential}\label{sec: discussion of effective potential}

\begin{figure}
\includegraphics[width=8.5cm]{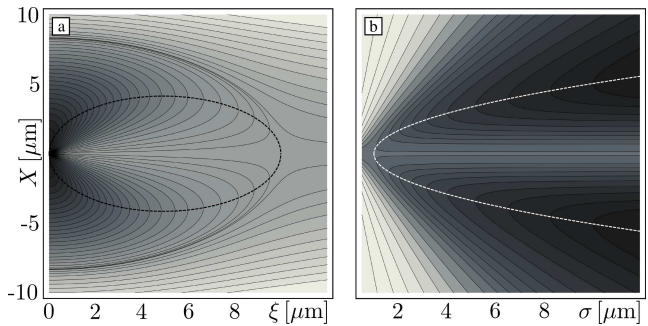}  
 \caption{(a) Contour plot of the effective potential for $Y=0$ for fixed $\sigma=10\,\mu$m as a function of $\xi$. With increasing $\xi$ the barrier gets lower and broader. (b) Same potential for fixed $\xi=1\,\mu$m and as a function of $\sigma$. For $\sigma>\xi$, the shape of the barrier close to the origin is almost conserved. However, the position of the minima increases with increasing $\sigma$. In both subfigures, the dashed line indicates the positions of the minima.}
  \label{fig: pots}
\end{figure}

The analytical prediction of the effective potential $V_\text{eff}$ for the $m_F=0$ component allows us to deduce its basic properties by a simple analysis of Eq.~(\ref{eq: V_eff}). The effects of the magnetic field creating the double well potential described in the previous section, are included in the second term in Eq. ~(\ref{eq: V_eff}). In order to obtain the double well structure this term needs to be negative. This is the case for laser light which is red detuned with respect to the $F=2$ state  and blue detuned with respect to the $F=1$ state, i.e, $0>\Delta>-\Delta_{\text{hfs}}$. An interesting case occurs for $\Delta=-\Delta_{\text{hfs}}/4$. Then the offset potential $V_0$ vanishes and the height of the barrier is equal to the maximal depth of the potential. For $0>\Delta>-\Delta_{\text{hfs}}/4$ the barrier-height is larger than the depth of the wells with respect to the continuum and for $\Delta_{\text{hfs}}<\Delta<-\Delta_{\text{hfs}}/4$ the height of the barrier is smaller than the depth of the wells with respect to the continuum.\\
In the following, we restrict our investigations again to the case $\Delta=-\Delta_{\text{hfs}}/2$. Since the double well only occurs for $0>\Delta>-\Delta_{\text{hfs}}$, this choice leads to a maximal detuning with respect to the $F=1$ and $F=2$ states. As expected from the numerical solutions provided in Sec.~(\ref{sec:numerical}), $V_\text{eff}$ shows a double well structure that is centered at the origin. The positions $(X_0,Y_0)$ of the two local minima are given by 
\begin{equation}\label{xmin}
 X_{0}=\pm \frac{\xi}{2}\sqrt{\sqrt{1+8\frac{\sigma^2}{\xi^2}}-3}
\end{equation}
and $Y_0=0$. It is obvious from Eq.~(\ref{xmin}) that the double well only exists if the discriminant is positive, giving rise to the condition $\xi<\sigma$.
For $\xi>\sigma$ one finds a single well potential, whereas for decreasing $\xi< \sigma$ the double well starts to build up. Starting in the limit $\xi\rightarrow0$ from a double well with a barrier with finite height but width going to zero, the distance between the minima increases with $\xi$ up to a local maximum $\Delta X^{\text{max}}_0=2(\sqrt{2}-1)\sigma$ at $\xi_{\text{max}}=\sqrt{3\sqrt{2}-4}\sigma$ and decreases for even larger $\xi$ up to $\xi_{\text{cr}}=\sigma$ where the minima vanish, thus transforming the energy surface to a single well potential. This behavior is illustrated in Fig.~\ref{fig: pots}(a) where a contour plot of the effective potential along the $X$-axis as function of $\xi$ is shown for a fixed width $\sigma=10\,\mu$m of the lasers; the dashed line indicates the positions of the minima.
Figure \ref{fig: pots}(b) shows a similar contour plot of the effective potential along the $X$-axis, but now as a function of $\sigma$ for fixed $\xi=1\,\mu$m. In this case, the shape of the barrier close to the origin is approximately conserved but the position of the minima increases with increasing $\sigma$.

\begin{figure}
\includegraphics[width=8cm]{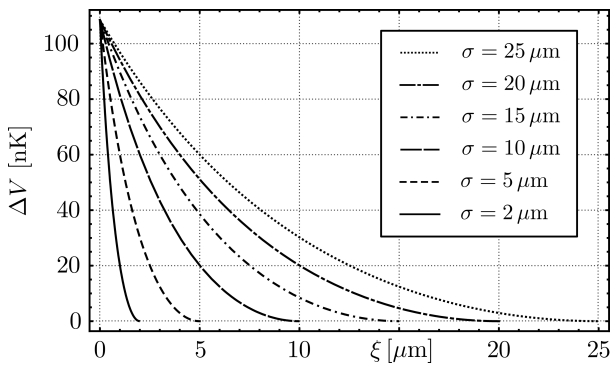}  
   \caption{Dependence of the height of the barrier $\Delta V $ on $\xi$ for different $\sigma$. For increasing $\xi$ the height of the barrier decreases monotonically.}
  \label{fig: fig8}
\end{figure}
The height of the barrier is given by $\Delta V = V_\text{eff}(0,0)-V_\text{eff}(X_0,0)$ and is directly proportional to the intensity of the lasers. Figure \ref{fig: fig8} shows the dependence of $\Delta V$ on the parameter $\xi$ for different values of $\sigma$ and fixed laser intensity $I=10$ W/m$^2$. The behavior of $\Delta V$ for different $\sigma$ is qualitatively very similar: the height of the barrier decreases with increasing $\xi$ monotonically. For fixed $\xi$, a more narrow laser entails a more shallow double well. Since the range of $\xi$ is determined by the condition $\xi<\sigma$, a more narrow laser necessitates furthermore tighter magnetic traps, i.e., smaller $\xi$. 
Note that the positions of the minima do not depend on the intensity of the lasers. Thus, by increasing the laser intensities one can increase the height of the barrier without changing the position of the minima. In this way, the number of trapped states in each well can be controlled. 

%
\subsection{Effective Potential for a Single Laser}
One can use the same semi-analytical procedure as above in order to predict the effective potential for a single laser. One finds
\begin{equation}\label{eq: V^1_eff}
 V^1_{\text{eff}}=\frac{1}{2}V_0+ \frac{1}{8}\left(1-\frac{\xi^2}{\xi^2+R^2}\right)V_\infty 
\end{equation}
where $R^2=X^2+Y^2$. The comparison of the effective potential (\ref{eq: V^1_eff}) and the corresponding numerical solutions shows again an excellent agreement. As opposed to the Raman setup involving two lasers, in the case of a single laser the resulting trapping potential for the $m_F=0$ component is rotationally symmetric. The ring-shaped minimum is located at 
\begin{equation}
 R_{0}=\sqrt{-\frac{5}{6}\xi^2+ \frac{1}{6} \xi^2\sqrt{1+12\sigma^2\xi^2}}
\end{equation}
and exists for $\sigma>\sqrt{2}\xi$. Starting at a barrier with arbitrarily small width with maximal height for $\xi\rightarrow0$, the barrier at the origin gets lower as $\xi$ increases and eventually vanishes as $\xi=\sigma/\sqrt{2}$. The distance of the minimum to the origin increases with $\xi$ up to the local maximum at $\xi_{\text{max}}=\sqrt{5\sqrt{3/2}-6} \sigma$ with a value $R_{\text{max}}=(\sqrt{3}-\sqrt{2})\sigma$. Then it decreases again and becomes zero at $\xi_{\text{cr}}=1/\sqrt{2}\sigma$ thereby transforming the potential into a single well. Note that the potential is not a simple superposition of a magnetic single well potential and a repulsive potential created by the laser. The spatial structure is a direct consequence of the spatially dependent light shift potential of the laser, giving rise to a barrier which is smaller than the width of the laser.

%
\section{Loss Mechanisms}
In the previous sections we have shown that for appropriate parameters the discussed combination of external fields leads to a confinement for two components of the ground state manifold.
 Let us  discuss in this section possible loss mechanism for these potentials.
%
\subsection{Lifetime of the Intermediate State}
One loss channel results from the coupling of the ground manifold of states to the excited $5P_{1/2}$ states. Despite the fact that the Raman lasers are detuned with respect to the excited state, there is a finite probability to excite the atom to this state due to the width of the state and the width of the lasers. The excited atom can subsequently decay spontaneously to the untrapped ground state. The resulting lifetime of the dressed state can be estimated by applying perturbation theory, leading to $\tau_\text{eff}=\tau (\Delta/ \omega_{\text{cp}})^2$ with $\tau$ being the lifetime of the unperturbed excited state and $\omega_{\text{cp}}$ the coupling matrix element of the ground state to the excited state. For our parameters ($\Delta=-\Delta_{\text{hfs}}/2$, $I_1=I_2=10$ W/m$^2$, and $\tau=27$ ns) we get an effective lifetime of $\tau_\text{eff}\sim 27$ ms. The latter can be increased by decreasing the intensities of the lasers. However, one has to bear in mind that this will reduce the depth of the trapping potentials as well. 

\subsection{Inelastic Collisions}
Another loss mechanism occurs if more than one atom is loaded into the potential due to the mutual interaction of the atoms. 
This mechanism can be estimated on a mean field level, incorporating an effective coupling coefficient that determines the interaction between the atoms. In the field-free case one gets a population transfer from one component to another due to interaction when there is an overlap between the wave functions of two components.
For our setup, however, one obtains new dressed states that are superpositions of the field-free states. Hence, one obtains a state changing contribution due to interaction even if only one dressed state is occupied.  However, for the above discussed parameters this additional term can be neglected.



\section{Conclusions and Physical Applications}
We investigated the trapping potentials for $^{87}$Rb ground state atoms simultaneously
exposed to a magnetic trap in a Ioffe-Pritchard like configuration and an 
optical trap in a Raman setup. The Raman lasers were detuned between the two 
excited $5P_{1/2}$ hyperfine states, the $F=1$ and the $F=2$ state.
By varying the offset field of the Ioffe-Pritchard trap, we demonstrated that the trapping potential of the $m_F=0$ component can be tuned from a rotationally symmetric single well to a double well trap; in the intermediate regime, one finds a cigar shaped trapping potential.
By applying a phase difference between the two Raman lasers, the resulting trapping potentials can be rotated about the propagation direction of the laser beams.
A semi-analytical formula for the potential surfaces has been derived. All relevant properties of the double well potential have been determined analytically as a function of the various trap parameters.
For a single excitation laser, the proposed scheme results in a ring-shaped trap for the $m_F=0$ component.

In order to exploit the unique features of the above discussed potentials, one can think of various experiments. For example, one might trap the $m_F=0$ component in a 
double well potential which gives for the $m_F=-1$ component a single well potential located at the center of the barrier. For an asymmetric occupation of the wells one can then observe tunneling of atoms trapped in the double well potential through the atoms trapped in the single well potential. The oscillation frequency of the tunneling can thus be investigated as a function of the occupation number of the second component located at the barrier, which is reminiscent of a single atom transistor \cite{Micheli}. We performed corresponding numerical simulations of the Spinor Gross Pitaevskii equation and observed indeed an increase of the oscillation period with increasing occupation number of the second component. 

In a similar setup as mentioned above 
(one component is exposed to a a double well potential with a narrow barrier and one component to single well potential centered at the position of the barrier) one might trap atoms in the component exposed to the single well potential and then transfer all atoms by an rf-pulse in the ``double-well'' component that feels a sharp potential maximum at this point. Depending on the energy of the atoms the condensate wave function would consequently either split into two parts or the whole wave function would move into one direction, which could be used as a test of the validity of the Gross Pitaevskii equation \cite{Streltsov07}. Moreover, the possibility to rotate an anisotropic single well trap (which arises for the $m_F=-1$ component) opens up a new possibility to study superfluids under rotation \cite{rev_fetter,PGK:MPLB:04,review}. The possibility to rotate the double well potential allows one to create an effective ring potential by rotating the potential fast enough so that the atoms feel a time averaged potential \cite{Lesanovsky07}. In this way the transition of a double well to a ring-shaped potential can be investigated. Moreover, one can exploit the feature that one can drive the potential surface of the $m_F=0$ component by modulating the Ioffe field strength to investigate non-equilibrium driven systems \cite{lenz08,lenz09}.

As an extension to the present work, it would be interesting to extend the studies to different propagation directions of the laser beams with respect to the orientation of the magnetic trap. Furthermore, one can vary the shape of one or both laser beams by using excited modes.

\end{document}